\documentclass[runningheads]{llncs}
\usepackage[T1]{fontenc}

\usepackage{geometry}
\usepackage{amsfonts}
\usepackage{amsmath}
\usepackage{graphicx}
\usepackage{tikz}
\usepackage{enumitem}
\usepackage{algorithm} 
\usepackage{algpseudocode}
\usetikzlibrary{positioning}
\usetikzlibrary{shapes.geometric}
\usepackage{changepage}

\usepackage{pgfplots}
\pgfplotsset{width=10cm,compat=1.9}

\begin{document}
\title{Functional Encryption in Secure Neural Network Training: Data Leakage and Practical Mitigations}
\titlerunning{FE in NN: Data Leakage and Mitigations}

\author{Alexandru Ioniță\inst{1}\orcidID{0000-0002-9876-6121} \and
Andreea Ioniță\inst{1}\orcidID{0009-0009-0007-3051}}
\authorrunning{Ioniță A., Ioniță A.}
%
\institute{Faculty of Computer Science, Alexandru Ioan Cuza University of Iasi, Iasi, Romania\\
\url{https://www.info.uaic.ro/}}

\maketitle              

\begin{abstract}
With the increased interest in artificial intelligence, Machine Learning as a Service provides the infrastructure in the Cloud for easy training, testing, and deploying models. However, these systems have a major privacy issue: uploading sensitive data to the Cloud, especially during training. Therefore, achieving secure Neural Network training has been on many researchers' minds lately. More and more solutions for this problem are built around a main pillar: Functional Encryption (FE). Although these approaches are very interesting and offer a new perspective on ML training over encrypted data, some vulnerabilities do not seem to be taken into consideration. In our paper, we present an attack on neural networks that uses FE for secure training over encrypted data. Our approach uses linear programming to reconstruct the original input, unveiling the previous security promises. To address the attack, we propose two solutions for secure training and inference that involve the client during the computation phase. One approach ensures security without relying on encryption, while the other uses function-hiding inner-product techniques.
\end{abstract}

\begin{adjustwidth}{2cm}{2cm}
\small Accepted to the 27th International Symposium on Research in Attacks, Intrusions and Defenses (RAID), 2025. © 2025 IEEE.
The final version will be published by IEEE and available via IEEE Xplore.
\end{adjustwidth}

\section{Introduction}
With the rapid growth of interest in machine learning and
artificial intelligence, more and more services are outsourced
to cloud systems to leverage higher computational power. In
this context, Machine Learning as a Service (MLaaS) emerges
as a practical solution, offering users access to ready-made
tools and scalable resources for developing, training, and
deploying models directly in the cloud. A major security
concern arises during the training phase, where the cloud
provider may gain access to the full training dataset. This
is a considerable impediment for machine learning models
that operate on sensitive data. An obvious example could be
medical information that should not be leaked to third parties.
This is usually achieved through anonymization but it does not
guarantee protection against inference attacks. Other use cases
involve company-specific information or pieces of intelligence.
In both cases, a method for encrypting the data and training
the model on the encrypted data would be of great interest

Modern cryptography offers vast options for encryption expressivity and granular access control. Starting with identity-based \cite{boneh2001identity} and attribute-based encryption \cite{goyal2006attribute}, which provide granular access over encrypted data. On the other side, \emph{functional encryption} (FE) has become very popular in recent years. This encryption primitive can permit only partial decryptions of the ciphertext: a decryption key has a function attached to it, and it is allowed only to decrypt the result over a function applied to the initial plaintext. 

FE has been used lately in a neural network environment to provide the ability to train over encrypted data \cite{xu2019cryptonn,panzade2023fenet}. Their approach provides a general framework to transform almost any neural network into one that could train over encrypted data. The computation overhead they provide, alongside the accuracy of the modified network compared to the initial one,
indicate that these systems could be used in practice. 
However, an aspect that has not been taken into account is the information leakage of these systems in the training phase. These articles state that \emph{inference} attacks, such as \cite{ligier2017information,carpov2020illuminating}, where a large amount of data is collected to retrieve information about training data, are outside their scope. However, attacks that directly try to obtain the input features from the partial training data are omitted.

\subsection*{Contribution}

We address the security issue of using FE to achieve secure training on encrypted data. We show that using the current proposed neural
network constructions from FE is not safe, since they recover the secret immediately before computing the activation function for the first hidden layer. 

We stress that our attack follows a novel approach compared to previous attacks on machine learning using FE. We make use of only the information leaked in the intermediate values in the training process without needing to collect information
from more than one sample, as most other attacks do. 

We provide practical and efficient results of our attack being applied to two neural networks trained on the CIFAR-10 \cite{cifar} dataset.

Finally, we propose two solutions to mitigate the attack: MITIG1 which utilizes function-hiding inner product encryption (FHIPE), while the other (MITIG2) avoids cryptographic methods entirely but relies on the client's participation in the computation process. We provide an implementation for MITIG2 over the MNIST \cite{MNIST} dataset, and compare it to previous secure training solutions. 


\section{Preliminaries}

\subsection{Functional Encryption}

FE is a type of public-key encryption that provides partial decryption, which means it can provide the result of a function on the encrypted data.
More formally, a FE scheme is composed of four algorithms, as follows:

\begin{description}
    \item[setup($\lambda$)]: The setup algorithm receives a security parameter $\lambda$ and returns the public and private keys, $mpk$ and $msk$.
    \item[encrypt($X$, $mpk$)]: The encryption algorithm receives an input $X$, and encrypts it under the public key $mpk$, returning a ciphertext $ct$.
    \item[keygen($f$, $msk$)]: The key generation algorithm receives a function $f$ and the secret key $msk$, and returns a decryption key $sk_f$.
    \item[decrypt($ct$, $sk_f$)]: The decryption algorithm recovers $f(X)$ from the decryption key $sk_f$ and the ciphertext $ct$ (which is obtained by encrypting the original message $X$)
\end{description}

The neural network training models we will study use different flavors of \emph{Inner Product} FE. In short, the main functionality needed is to compute the inner product $\langle x,y\rangle$ between an encrypted vector $x$ and a vector $y$ associated with a decryption key. This operation is essential in the context of neural networks, where the computation of inner products between input data and weight vectors is a fundamental step repeated at each layer during the forward pass.

Let \( W \in \mathbb{R}^{m \times n} \) be a real-valued matrix, and let \( X \in \mathbb{R}^n \) be a real-valued column vector. 
The inner product $Z_1$ of \(\langle W,X\rangle \in \mathbb{R}^m \) is defined as:

\[ \begin{pmatrix}
    x_1 \cdot w_{11} + x_2 \cdot w_{12} + \dots + x_{n} \cdot w_{1n}\\
    x_1 \cdot w_{21} + x_2 \cdot w_{22} + \dots + x_{n} \cdot w_{2n}\\
    x_1 \cdot w_{31} + x_2 \cdot w_{32} + \dots + x_{n} \cdot w_{3n}\\
    \vdots \\
    x_1 \cdot w_{m1} + x_2 \cdot w_{m2} + \dots + x_{n} \cdot w_{mn}
\end{pmatrix}=
\begin{pmatrix}
    z_{11}\\
    z_{12}\\
    z_{13}\\
    \vdots\\
    z_{1m}
\end{pmatrix}
\]

A first FE scheme with this functionality called FEIP (FE for Inner Product) was proposed in \cite{abdalla2015simple}. \cite{kim2018function} proposed a FE scheme for inner product called \textbf{FHIPE} with function hiding capabilities, meaning that the vector $y$ remains hidden to the holder of the decryption key. However, this comes with a considerable overhead in encryption and decryption due to the use of bilinear pairings in algebraic groups.

 To successfully decrypt any of the aforementioned schemes, it is necessary to compute the discrete logarithm, a requirement that introduces substantial computational overhead to the entire process. Although other instantiations of FE could be made, for example by using Lattices under the Learning with Errors assumption \cite{MKMS2022}, they also imply a considerable computational overhead.

As our attack is agnostic to the underlying mathematical construction of the FE schemes, we omit the full technical details here, but they can be consulted in the Appendix \ref{app:fe}.

\subsection{Linear Programming}
\label{app:lp}
Linear programming is the process of minimizing a linear objective function subject to a finite number of linear equality and inequality constraints. \cite{karloff2008linear}\newline
 For example, linear programming would find a solution to maximize or minimize 
 \begin{equation} 
    P=p_1x_1 + p_2x_2+\dots p_kx_k \text{, having the following constraints:}
\end{equation}
 \begin{align*}
     a_{11}x_1 + a_{12}x_2+\dots a_{1k}x_k\leq q_1\\
     a_{21}x_1 + a_{22}x_2+\dots a_{2k}x_k\leq q_2\\
     \vdots \\
     a_{n1}x_1 + a_{n2}x_2+\dots a_{nk}x_k\leq q_n\\
     x_1, x_2, \dots,x_k\geq 0
 \end{align*} 

\subsection{Adversarial Model}
As we are discussing a security problem, it is important to review the existent types of adversaries. In this way, we can decide in which adversarial model the security of a system is met.\newline
\textbf{Semi-honest model.} This adversary participates correctly in the protocol, but collects information so that he can learn as much as possible from it.\newline
\textbf{Security in the semi-honest model.} A secure protocol in a semi-honest setting has the property that the view obtained in the execution by an adversary is the same as the view obtained in the ideal model. In other words, the attacker should not be
able to gain any additional information beyond their authorized
output by applying any polynomial-time algorithm to the data
available to them.
In our model, the CSP acts as a semi-honest adversary: it
follows the protocol as specified but aims to infer additional
information from any data it can observe during the compu-
tation.

\subsection{Notations and symbols used}

\begin{tabular}{c l}
    \textbf{Symbol} & \textbf{Description} \\
    $X$ & Input layer \\
    $[[X]]$ & Encryption of $X$ \\
    $n$ & Size of the input $X$ \\
    $W$ & Weight matrix of size $m \times n$ \\
    $Z_1$ & Result of inner product $\langle W, X \rangle$ \\
    $\sigma$ & Activation function \\
    $b$ & Bias vector \\
    $A_1$ & First hidden layer, $A_1 = \sigma(Z_1 + b)$ \\
    $m$ & Size of the first hidden layer \\
    $dZ_1$ & Derivative of $Z_1$ \\
    $sk_f(W)$ & Functional decryption key of $W$ on $f$ function\\
    $p_{i,j}$ & Pixel value at position $\{i,j\}$ \\
\end{tabular}

\section{Related work}

\subsection{NN using FE}

With the growing demand for security, an increasing number of cryptographic schemes are being integrated into neural networks. As highlighted in \cite{PaTa2023Privacy}, several machine learning models support inference using functional encryption (FE), but only one—CryptoNN \cite{xu2019cryptonn}—also enables private training over encrypted data. More recently, we have identified a second such model, FeNet \cite{panzade2023fenet}, which extends the ideas introduced by CryptoNN.

\subsubsection{CryptoNN}
\begin{figure}
    \centering
\includegraphics[scale = 0.19]{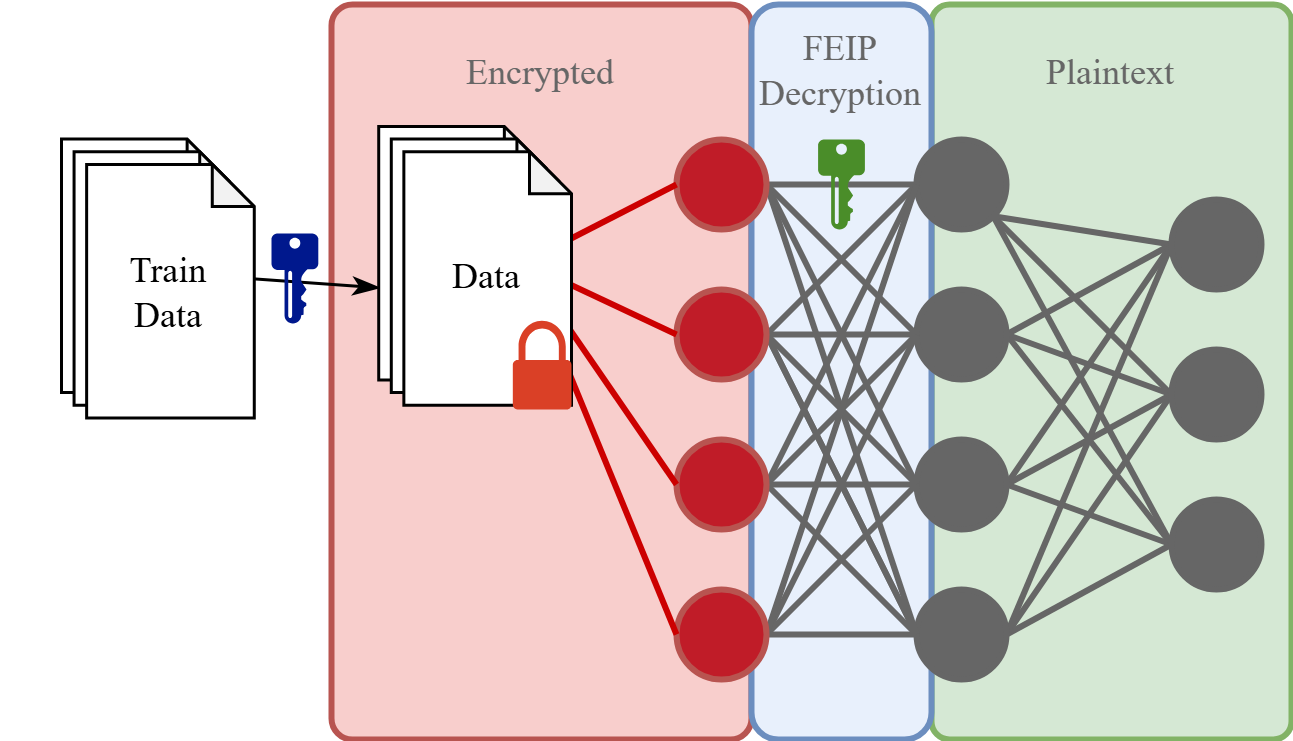}
    \caption{CryptoNN feed-forward architecture. The client encrypts the input data $X$ using functional encryption (FE) and sends the ciphertext $[[X]]$ to the CSP. In the first hidden layer, the CSP uses the functional decryption key $sk_f(W)$ to compute the inner product $\langle W, X \rangle$ in plaintext. From this point onward, the rest of the neural network operates on plaintext data.}
    \label{fig:1}
\end{figure}

In 2019, \cite{xu2019cryptonn} proposed the use of FE to train neural networks directly over encrypted data. They make use of the previously presented FE scheme FEIP\cite{abdalla2015simple} in order to encrypt only the training data set. The authors calculate the inner product of the encrypted data and obtained the result in plaintext in the first hidden layer. The rest of the network is in plaintext, except for the last layer which is encrypted using FEBO. The complete scheme can be seen in Figure \ref{fig:1}.

 They propose a framework, called CryptoNN, that has accuracy comparable to that of the original neural network model. Essentially, CryptoNN inserts a secure feed-forward step and a secure back-propagation and evaluation steps into the training phase of the normal neural network. 
A central authority is responsible for generating the cryptographic parameters, including public and private keys, as well as function-specific decryption keys for the server to decrypt the function result.\newline 

In order to support matrix computation over encrypted data, they construct secure matrix computation using FE for inner product and for basic operations. Their proposal has three parts: the encryption of input, key generation for the permitted set of functions and finally, the secure computation of function. The first hidden layer takes each encrypted sample $X_i$ and multiplies it with the weights, calculating $\langle W X\rangle$. Since the input is encrypted with the FE scheme, this process becomes:
\begin{equation}
    Z_1 = sk_f(W)\cdot [[X]] = \langle W,X\rangle.
\end{equation}
From that point onward, the standard feed-forward process continues on unencrypted data. It takes the output of the first hidden layer, which is computed in plaintext after the decryption and continues throughout the rest of the network.

It is worth mentioning that even if the first hidden layer is calculated using $[[X]]$, the result is in plaintext because of the properties of FE.

The security of the FE scheme is given by the \emph{Decisional Diffie-Hellman} assumption. However, they address some concerns about possible attacks after the decryption phase. 
Thus, not related to the security of the encryption scheme. They assume that the server is not an active attacker that will collect a “representative plaintext dataset” for the encrypted training data. 

In other words, they prove that an honest-but-curious attacker would, under no circumstances, be able to reconstruct anything with the data that it has. However, in the further sections, we propose an attack on FE neural network schemes, by using FE in order to find the input layer. In our attack, we do not collect a representative plaintext dataset, as we only need one sample in order for the attack to take place.

\subsubsection{FENet}
Although privacy-preserving machine learning (PPML) using FE is still in the early stages, another article that uses FE in order to train neural networks in a secure way is presented by Panzade et al. in \cite{panzade2023fenet}. They are the first to use a function hiding inner product encryption in PPML, but also provide a version of encryption of the data in NN using FEIP. 

The assumption is that all the computations are made in the MLaaS setting, as in their framework proposal they consider 3 entities: a key management authority, a cloud server, and a client. Similar to CryptoNN, the following security is honest but curious. 

The workflow of FENet is actually very similar to CryptoNN:
after obtaining master keys from KMA, the client encrypts the input using public keys. However, they use different public keys for each sample.  The ciphertexts are then sent to the CS for training, which generates a decryption key based on the weight vector and decrypts the result of the inner product between the input and the weights.

The described process is the first step in the forward propagation, which they refer to as \emph{secure forward propagation}. However, for secure backpropagation, the authors use the categorical cross-entropy loss function and stochastic gradient descent algorithm as an optimizer to adjust the weights. In the inference stage, things are very similar to the secure propagation phase. The sample must be encrypted (with the same encryption algorithm previously used) and after performing secure forward propagation, the classification result is returned to the client.

We stress that the FE is used very similarly in FENet as in CryptoNN in the forward propagation phase. It discloses the plaintext of the product between the
encrypted input features and the weights of the first hidden layer, which, as we will show later, leads to information leakage. Although FENet uses FHIPE, the server knows the plaintext values for the weights, thus making it susceptible to our attack, as we can see in Section \ref{sec:our_attack}.

\subsection{Attacks on Encrypted ML}
\label{sec:rw:attacs}

Right from the beginning of FE use in neural networks, some questions were raised regarding the real security of this solution. In 2020 Carpov et al. \cite{carpov2020illuminating} the authors examine whether current practical FE schemes leak information about the encrypted
input data during the classification process. They show that
using this leakage, neural networks can partially reconstruct
information that should remain confidential. More precisely,
these classifiers involve a first layer with a linear or quadratic
function realized using a FE scheme. The output of this layer
is then used as training data for the neural network, which acts
like an inversion model.

Melis et al. \cite{melis2019exploiting} explored unintended feature leakage in collaborative learning by showing that shared gradient updates can reveal sensitive attributes about the participants' local data. Their method involves training an auxiliary classifier to infer private properties from gradients without requiring access to the raw inputs. Although effective, their attack primarily targets feature-level leakage rather than full input reconstruction, and its success depends on the amount of leaked gradient information and the correlation between the target attribute and the model’s objective.

Zhu et al. \cite{ZhLH2019} introduced \emph{Deep Leakage from Gradients} (DLG), a new attack that demonstrates how gradient 
information alone can leak sensitive training data in federated learning. 
Their method reconstructs both the input and its label by optimizing a dummy input
such that its computed gradient matches the observed gradient sent by a client.
While their approach works well during the training process, where the weights 
are re-computed after each iteration, in the case of batched training, the attack becomes inefficient due to the aggregation of gradients.

Another approach to recovering information from inner-product FE-based data classification is presented in \cite{ligier2017information}. They present two attacks: one using an artificial neural network and the other making use of principal component analysis. The principal component analysis method is a statistical procedure that transforms a number of correlated variables into a smaller set of uncorrelated variables, still containing most of the information in it.  Both attacks are split into k parts, one for every class, so, this is one of the differences compared to our attack, as we do not need any information besides the weights and the result of the inner product. Their approach using neural networks needs a new neural network for each class in order to find the input layer, the fact that makes it effective but time and resource-consuming. 

Carlini et al. have another technique in terms of attacks \cite{carlini2019secret}. They discuss a type of leakage provoked by the memorization of sensitive data in the neural network. They propose a way of verifying if the dataset is prone to be the victim of a memorization attack.

\section{Our Attack}
\label{sec:our_attack}
\begin{figure}
    \centering
    \includegraphics[width=0.65\linewidth]{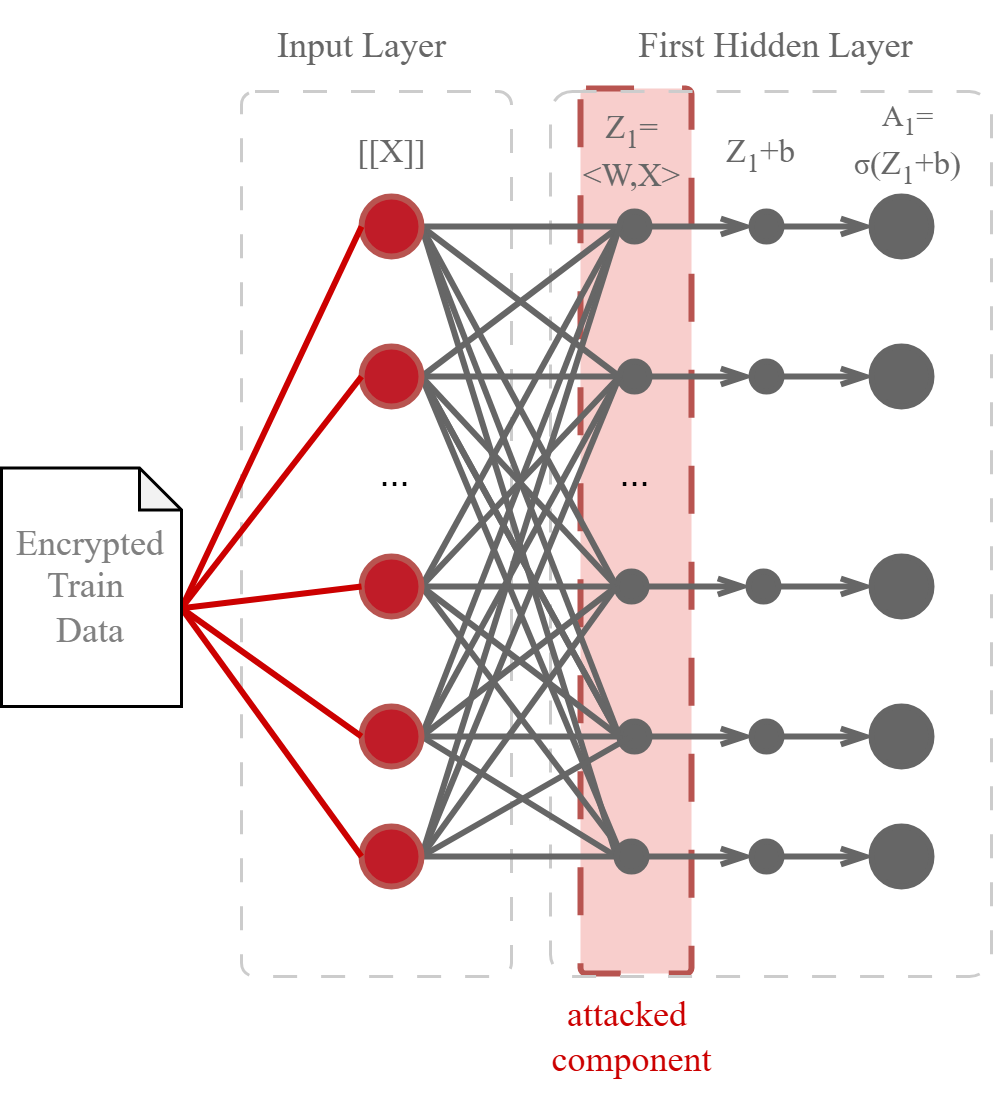}
        \caption{The figure illustrates the attack point in both training and inference phase: although the input $[[X]]$ is encrypted, the first hidden layer is in plaintext. Using the functional decryption key $sk_f(W)$, the inner product $Z_1 = \langle W, X \rangle$ is revealed, which is then added to the bias $b$ and passed through the activation function to compute $A_1 = \sigma(Z_1 + b)$.}
    \label{fig:attack_point}
\end{figure}

In this chapter, we present an attack applicable by a CSP
while supervising a training process. While in the training
phase, the input is encrypted using FE, we show that the
information leaked during the decryption in the first hidden
layer of the training process can lead to a partial or full
recovery of the training data. Our attack applies to both NN
models that use FE in the training process, 
namely  CryptoNN \cite{xu2019cryptonn} and FENet \cite{panzade2023fenet}.

 What do all these neural networks that we propose an attack for have in common? Firstly, the only layer that is encrypted is the first one, the input layer. This distinction is visually depicted in Figure \ref{fig:attack_point}, where the
input layer is shown as encrypted, while the first hidden layer
operates on plaintext data.The rest of the network, the hidden
layers use plain text data. Our attack uses the input of the
first hidden layer, which is recovered in plain, alongside the
weights, in order to reconstruct or learn information about the
encrypted input fed to the network.
 
We emphasize that the type of FE scheme used is irrelevant
to our attack, as it only relies on the plaintext revealed after
legitimate decryption. This also is within the boundaries of a
\emph{honest but curious} attacker.
The main idea of the attack is to find $X = [ x_1,\dots x_{n_1}]$ by making use of the first Layer, which is the dot product between the unknown $I$ and the known weights $W$. All this information is available to a Cloud administrator, which
provides the hardware support for training the network.

We now describe multiple scenarios in which the attack can
be applied, demonstrating its flexibility. All attacks variants
are constructed using a single sample in the training phase, to
emphasize that a large dataset is not needed to reconstruct the
input, in contrast with the attacks described at Section \ref{sec:rw:attacs}.

\subsection{Vulnerabilities when input layer is smaller than the first hidden layer}
\label{attack-large-layer1}

A straightforward attack scenario arises when the number
of neurons in the input layer is smaller than that of the first
hidden layer. This use case can be described as follows:

A straightforward attack scenario arises when the number of neurons in the input layer is smaller than that of the first hidden layer. This use case can be described as follows:
\begin{enumerate}
    \item $n$ is the number of neurons from the input layer $X$
    \item $m$ is the number of neurons before activation from the first hidden layer $Z_1 = [z_{11},z_{12},z_{13},\dots z_{1m}]$
    \item $n \leq m$
    \item $Z_1 = sk_f(W) \cdot [[X]]$ with the aid of \textit{FE} 
\end{enumerate}
By decrypting the weight matrix $W$ using the secret key corresponding to the inner product computation with the encrypted input $[[X]]$, we obtain the following plaintext result for $Z_1$:
\[ Z_1 = \begin{pmatrix}
    x_1 \cdot w_{11} + x_2 \cdot w_{12} + x_3 \cdot w_{13} + \dots + x_{n} \cdot w_{1n}\\
    x_1 \cdot w_{21} + x_2 \cdot w_{22} + x_3 \cdot w_{23} + \dots + x_{n} \cdot w_{2n}\\
    x_1 \cdot w_{31} + x_2 \cdot w_{32} + x_3 \cdot w_{33} + \dots + x_{n} \cdot w_{3n}\\
    \vdots \\
    x_1 \cdot w_{m1} + x_2 \cdot w_{m2} + x_3 \cdot w_{m3} + \dots + x_{n} \cdot w_{mn}
\end{pmatrix} \]

 In this matrix, we do not know $X$ but we can find it by solving a system of equations with Gaussian elimination: 

\[ \begin{pmatrix}
    w_{11} & w_{12} & w_{13} & \dots & w_{1n} \\
    w_{21} & w_{22} & w_{23} & \dots & w_{2n} \\
    w_{31} & w_{32} & w_{33} & \dots & w_{3n} \\
    \vdots & \vdots & \vdots & \dots  & \vdots \\
    w_{m1} & w_{m2} & w_{m3} & \dots & w_{mn}
\end{pmatrix} \cdot X = \begin{pmatrix}
    z_{11} \\
    z_{12} \\
    z_{13} \\
    \vdots \\
    z_{1m}
\end{pmatrix} \] 

This will lead to a full recovery of the hidden sample. Note that in order for this attack to take place we only need a single entry in the test dataset, and we can process each entry in the dataset individually.

\subsection{Attacking networks with small first hidden layer}

When the number of neurons in the first hidden layer is lower than the input size, then an attacker should not be able to fully recover the input, but it still should be able to discover some information about it. In order to better understand how big this leak is, we have conducted some practical tests against CryptoNN, but the same attack can also be applied to FENet, since also here the Cloud Server also knows the weights values.

One failed attempt using the \textit{Adam} optimizer is explained in the Appendix \ref{app:adam}.

On our second attempt, we have observed that \emph{Linear programming} fits like a glove for our problem. The set of constraints can be formed by both equations and inequalities, and we could limit the range in which the solution is searched. The only inconvenience is that, typically, a linear programming algorithm will aim to optimize the result
of some linear objective function, and this is not applicable in our attack. 

We have gradually developed three stages of our attack: simple linear programming with no augmentation, two versions where we tried to artificially augment the equations, and then inequalities we introduced in the system, based on the expectation of the input. Each of these stages is described below:

\subsubsection{Simple Linear Programming Attack}

\noindent
\bgroup
\setlength\tabcolsep{0.5pt}
\begin{table}
\caption{simple linear programming attack}
\label{fig:test-no-aug}
    \centering
    \begin{tabular}{|c|c|c|c|c|}
        \hline 
        Layer 1 & 500 & 750 & 900 & 1000  \\
        \hline
            \includegraphics[width=5em]{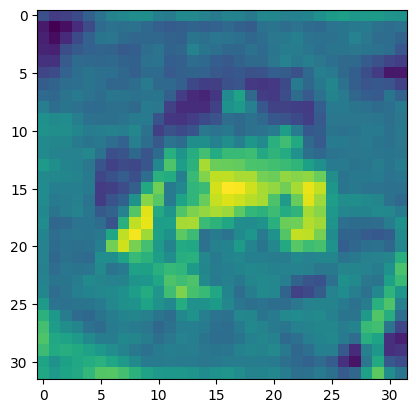} &
          {\includegraphics[width=5em]{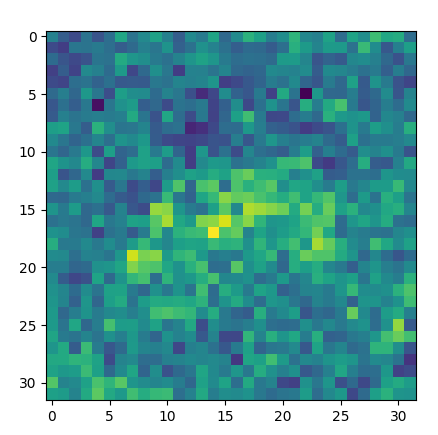} } &
          {\includegraphics[width=5em]{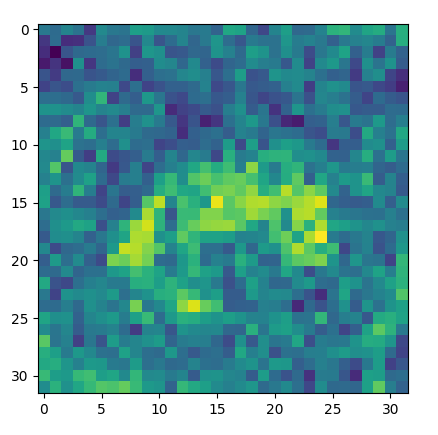} } &
          \includegraphics[width=5em]{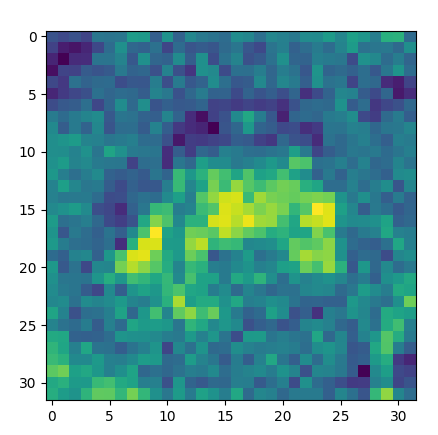} &
          {\includegraphics[width=5em]{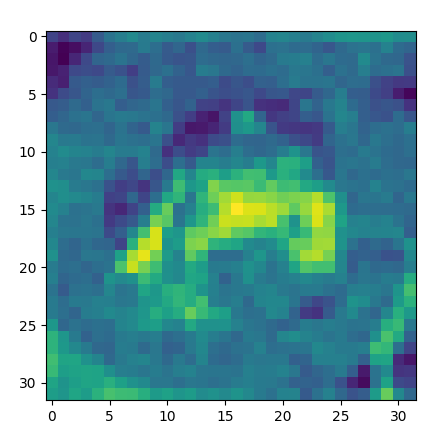} } \\
          \hline
    \end{tabular}
\end{table}
    \egroup

The first set of equations comes from the dot product $W \cdot X = Z_1$. 
In this
configuration, the only parameter we varied was the number
of neurons in the first hidden layer, but at the same time, this
number should have been less than the input layer($32 \times 32$).
So we tested with 500 neurons on the first hidden layer, then
750, 900 and 1000. The results can be seen in Table \ref{fig:test-no-aug}. In the
first column, we can see the original image, followed by the
recovered image in every one of the four test cases. As can
be seen, the area of interest (the frog) takes shape, but in a
very blurry and with a lot of noise in the first case, with 500
neurons. The increasing number of neurons makes the central
piece of the image stand out with more and more precision,
but the image still has a lot of noise. Only when the number
of neurons is very close to the input size (1024), the result
becomes accurate.

\begin{table}[hbt]
\caption{Test results for a neural network with a dense first layer}
\label{fig:attack-dense}
\hspace*{-2cm}
    \centering
    \begin{tabular}{|c||c|c|c|c|c|c|c|c|c|c|}
        \hline 
        Layer 1 & \multicolumn{3}{c|}{350}  & \multicolumn{3}{c|}{500}  & \multicolumn{3}{c|}{750}  \\
        \hline
        Ineq  & 1/16 & 1/9 & 1/4 & 1/16 & 1/9 & 1/4 & 1/16 & 1/9 &  1/4 \\
          \hline
            \includegraphics[width=5em]{photos/original/img0_original.png} &
            \includegraphics[width=5em]{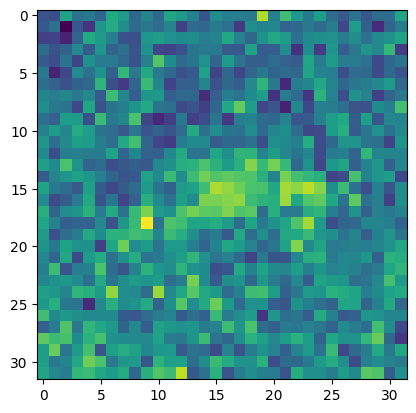} &
          \includegraphics[width=5em]{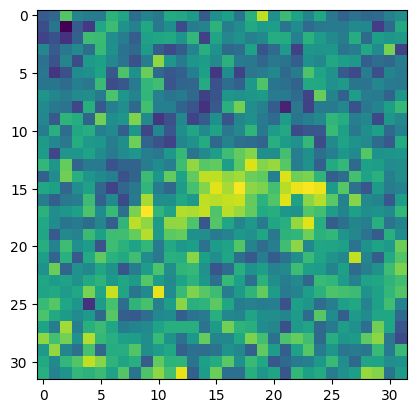} &
          \includegraphics[width=5em]{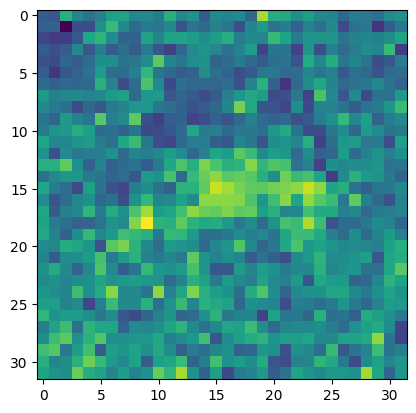} &
          {\includegraphics[width=5em]{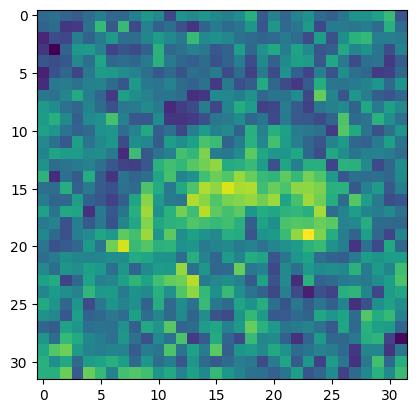} } &
          {\includegraphics[width=5em]{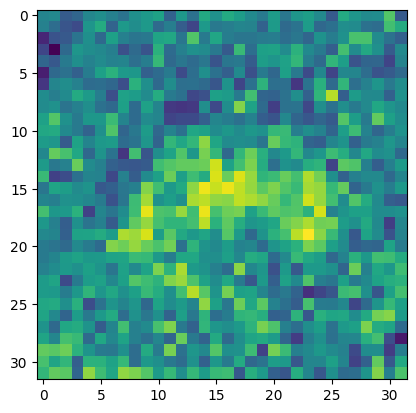} } &
          {\includegraphics[width=5em]{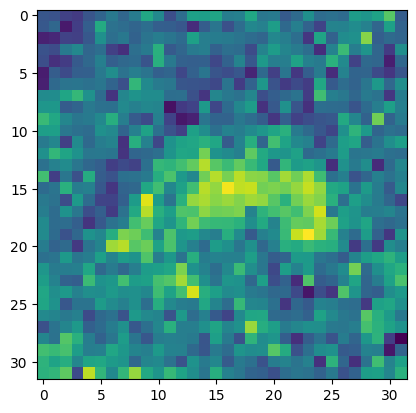} } &
          {\includegraphics[width=5em]{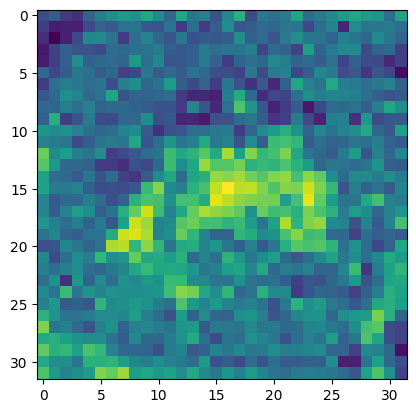} } &
          {\includegraphics[width=5em]{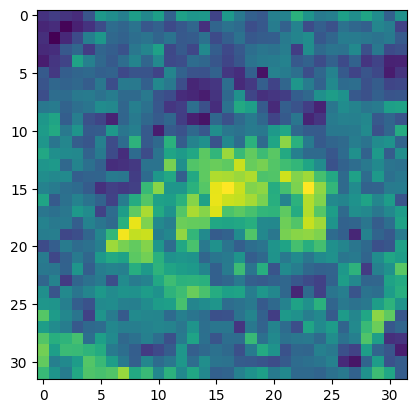} } &
          {\includegraphics[width=5em]{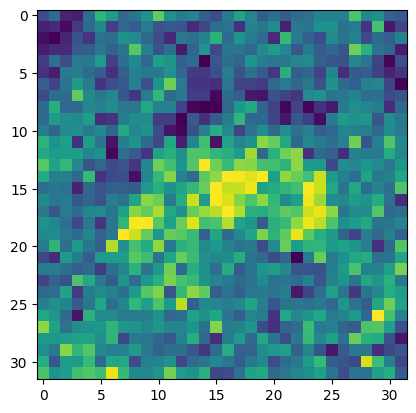} } \\
          MSE &
          0.018 &
          0.018 &
          0.016 &
          0.015 &
          0.014 &
          0.011 &
          0.008 &
          0.008 &
          0.010\\
          \hline
          \includegraphics[width=5em]{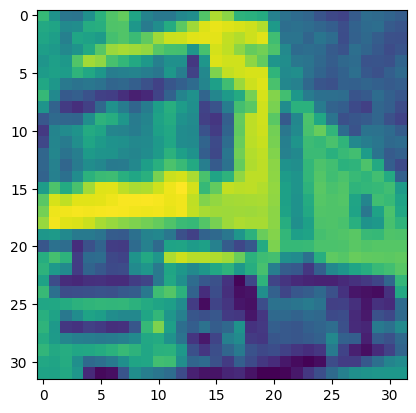} &
          {\includegraphics[width=5em]{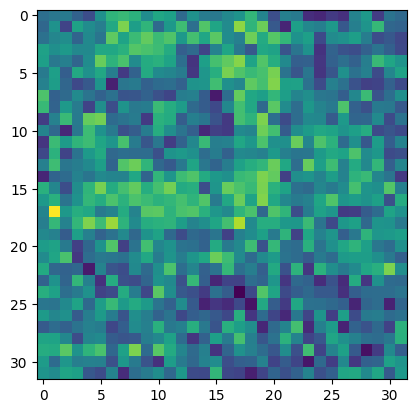} } &
          {\includegraphics[width=5em]{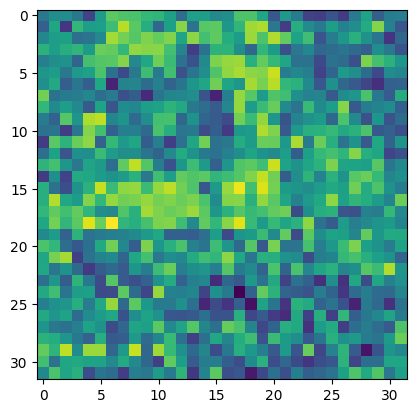} } &
           \includegraphics[width=5em]{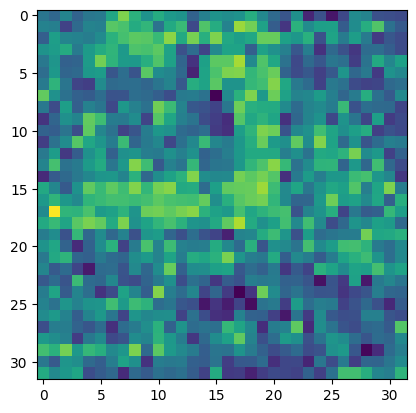} &
          {\includegraphics[width=5em]{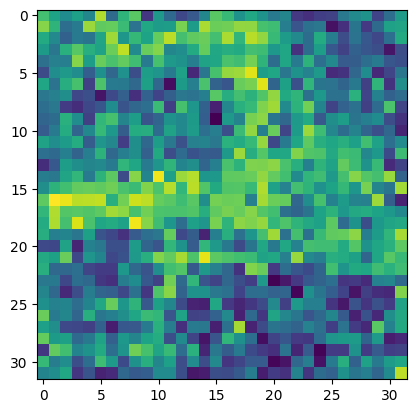} } &
          {\includegraphics[width=5em]{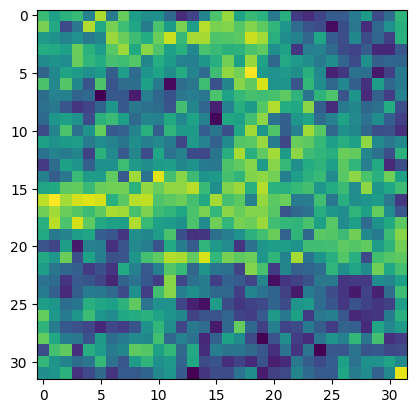} } &
          {\includegraphics[width=5em]{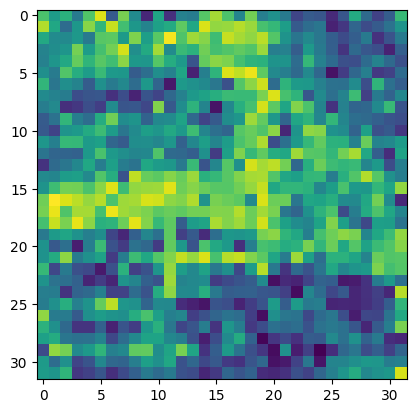} } &
          {\includegraphics[width=5em]{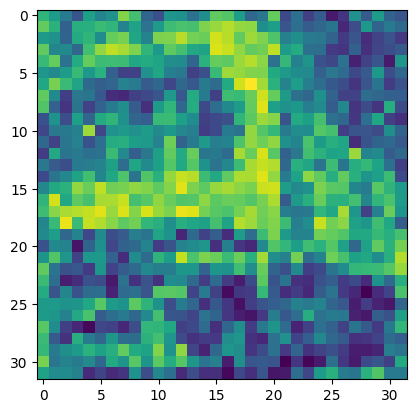} } &
          {\includegraphics[width=5em]{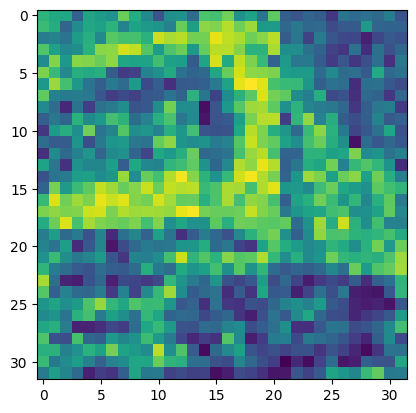} } &
          {\includegraphics[width=5em]{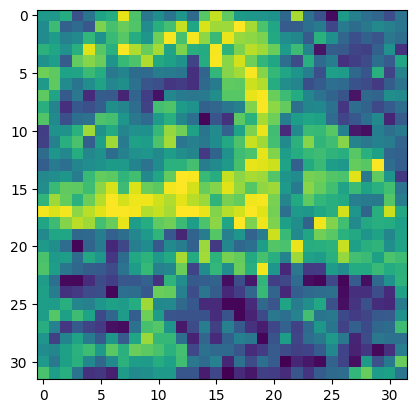} } \\
          \hline
          \includegraphics[width=5em]{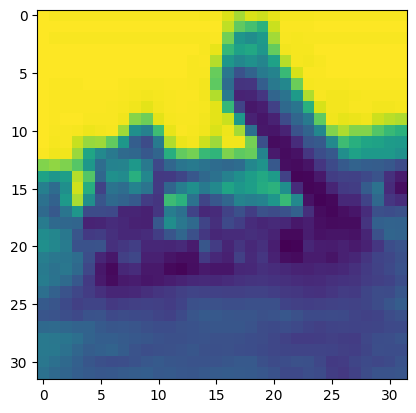} &
          {\includegraphics[width=5em]{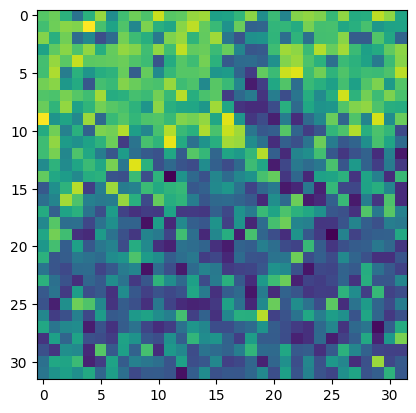} } &
          {\includegraphics[width=5em]{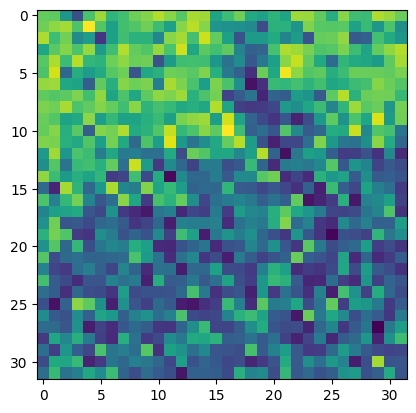} } &
           \includegraphics[width=5em]{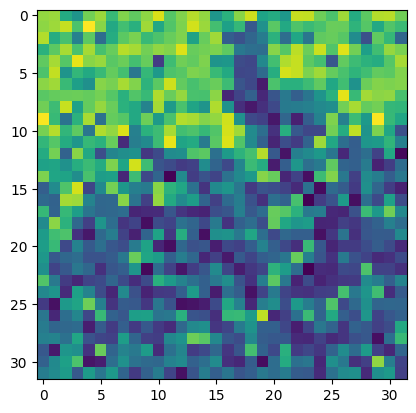} &
          {\includegraphics[width=5em]{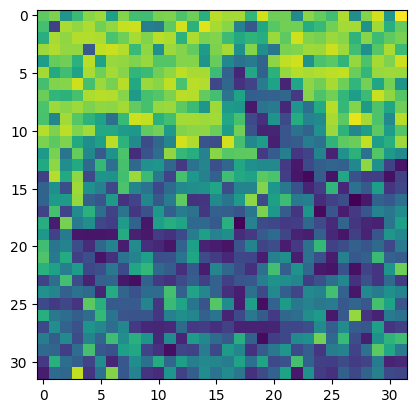} } &
          {\includegraphics[width=5em]{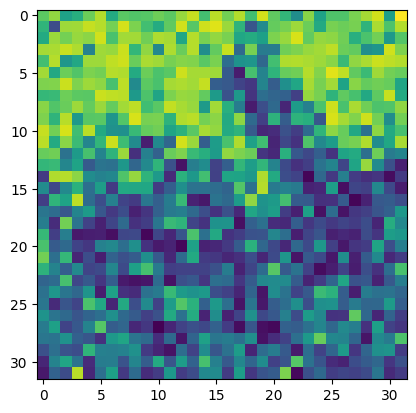} } &
          {\includegraphics[width=5em]{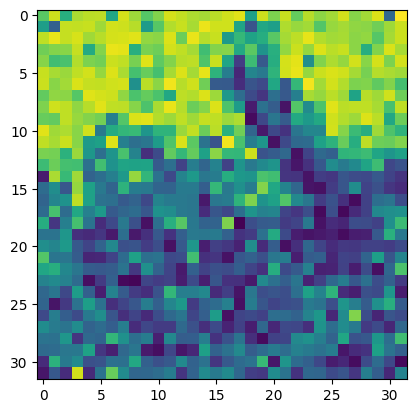} } &
          {\includegraphics[width=5em]{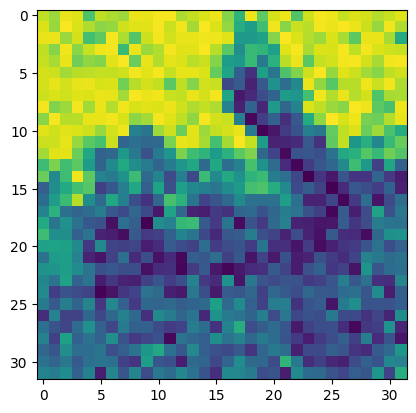} } &
          {\includegraphics[width=5em]{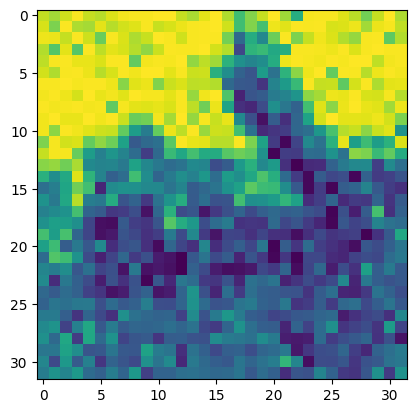} } &
          {\includegraphics[width=5em]{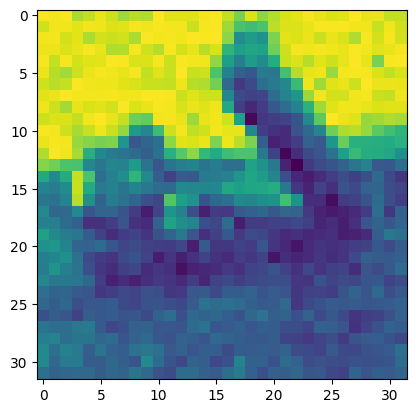} } \\
          \hline
            \includegraphics[width=5em]{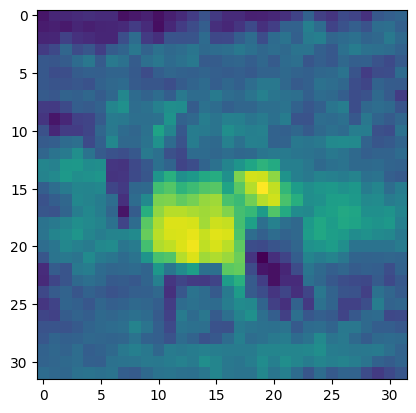} &
          {\includegraphics[width=5em]{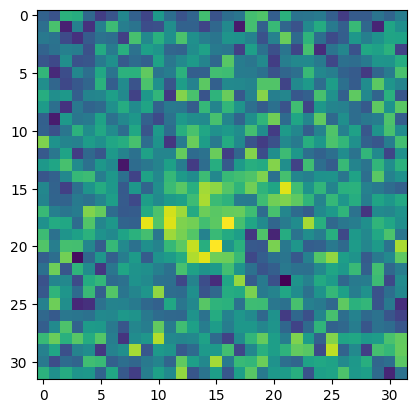} } &
          {\includegraphics[width=5em]{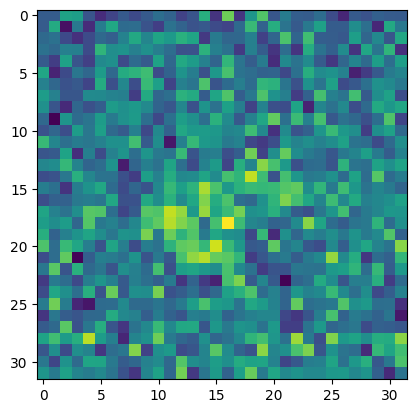} } &
           \includegraphics[width=5em]{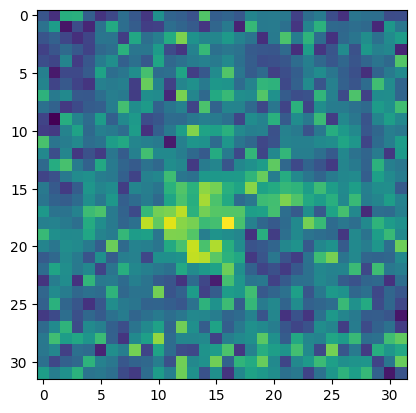} &
          {\includegraphics[width=5em]{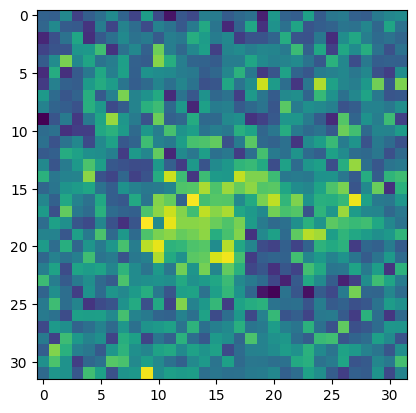} } &
          {\includegraphics[width=5em]{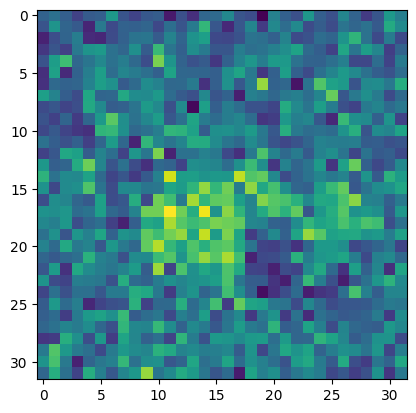} } &
          {\includegraphics[width=5em]{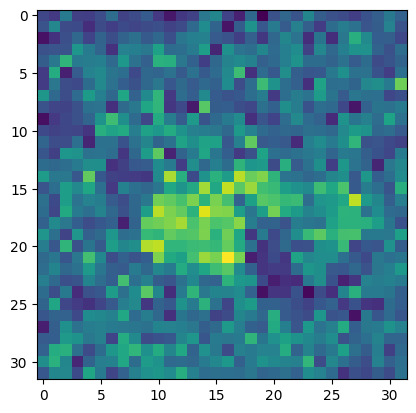} } &
          {\includegraphics[width=5em]{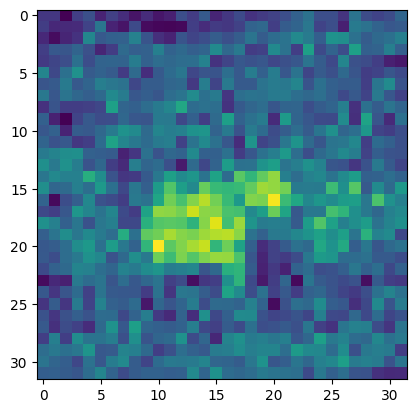} } &
          {\includegraphics[width=5em]{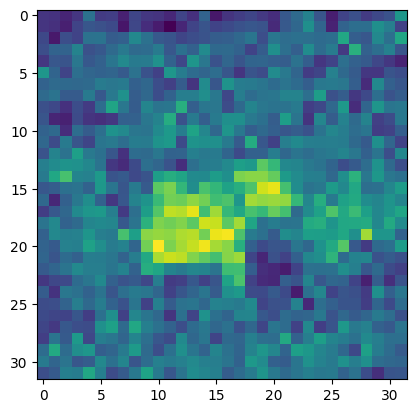} } &
          {\includegraphics[width=5em]{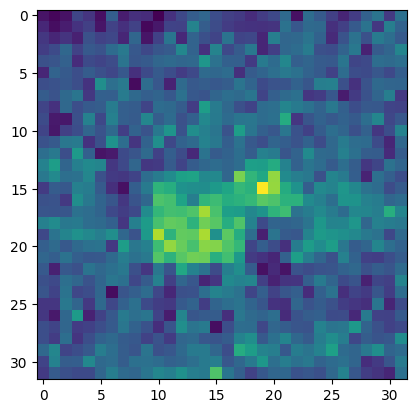} } \\
          \hline

    \end{tabular}
\hspace*{-1cm}
\end{table}

\subsubsection{Linear programming attack using augmentation equations}

The results were not anywhere near what we expected, reason why we added some other constraints, based on our empirical research. We observed that, in general, two consecutive pixels are of similar colors, which is why we decided to add equations in the form
\begin{equation}
    p_{i,j} - \frac{p_{i-1,j} + p_{i+1,j} + p_{i,j-1} + p_{i,j+1}}{4} =0. \label{eq:1}
\end{equation}


When it comes to stringent equations, we had two variables to play with: the number of neurons in the first layer and the number of pixels that are calculated with the formula indicated above. However, our goal was to increase the quality of the image by only adding more calculated pixels, not increasing the number of neurons, as we have already done in the previous case.

The augmentation with stringent equations was another failed attempt, as the quality of the image did not get any better as compared to the one with simple linear programming.

\subsubsection{Linear programming attack using threshold inequalities}

As augmentation with stringent equations was not of very much help, we turned our attention to adding some inequalities in order to solve the system, and therefore to increase the quality of the result image.

To have more flexibility and remove the imposed limitations in the system, we decided to transform the equation \eqref{eq:1} into two inequalities: 
\begin{equation}
         \left|p_{i,j} - \frac{p_{i-1,j} + p_{i+1,j} + p_{i,j-1} + p_{i,j+1}}{4} \right|\leq t \label{ineq:1}
\end{equation}

where $t$ is a chosen threshold. 

We have made tests with multiple instances of the neural
network, with different neuron count on the first hidden layer:
350, 500, 750 and 1000 neurons. Note that the rest of the
network, and even the backpropagation phase are irrelevant to
this attack. For each instance of the neural network we have
run five attacks, with different types of augmentation. First,
we have observed that for a neural network with a first dense
layer, the best results are obtained when the augmentation
inequalities are chosen in a systematic manner, rather than
random. Therefore, we have chosen 1/2, 1/4, 1/9 and 1/16
pixels, based on the value of i + j modulo 2, 4, 9 and 16.

For some of these tests we have visual and analitical results
in Table  \ref{fig:attack-dense}. In the first column, we can see the original image,
followed by nine tests, separated into three groups, by the
number of neurons in the first hidden layer. In each group
we have three additional classes, based on the number of
inequalities used (1/16, 1/9 and 1/4 out of the total number
of pixels). Under the visual representation, we have the Mean
Squared Error (MSE) between the original image and the
recovered one.

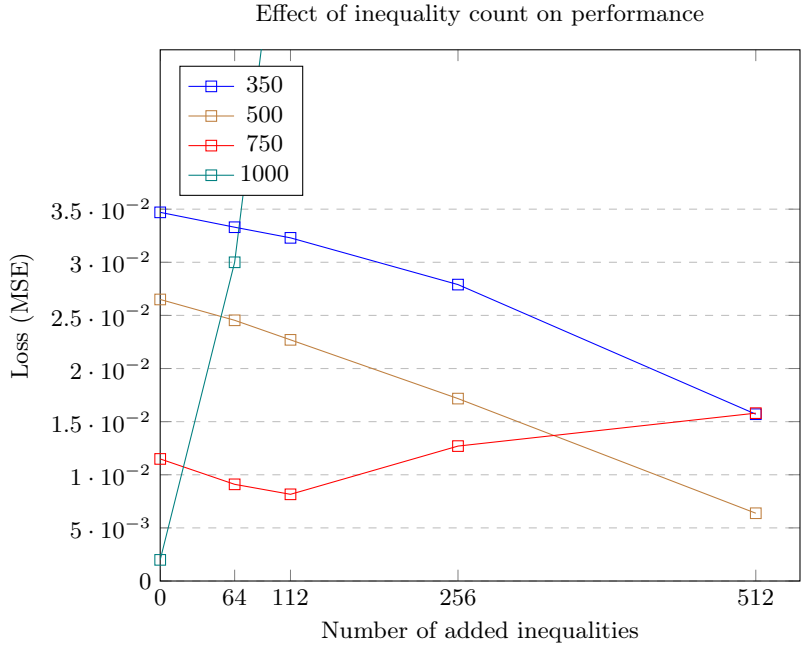
\begin{figure}
    \centering
    \begin{tikzpicture}
\begin{axis}[
    title={Effect of inequality count on performance},
    xlabel={Number of added inequalities},
    ylabel={Loss (MSE)},
    xmin=0, xmax=550,
    ymin=0, ymax=5e-2,
    xtick={0,64, 112, 256, 512},
    ytick={0,1.5e-2, 2e-2,1e-2, 2.5e-2, 3e-2, 3.5e-2, 5e-3},
    legend pos=north west,
    ymajorgrids=true,
    grid style=dashed,
    scaled ticks=false
]

\addplot[
    color=blue,
    mark=square,
    ]
    coordinates {
    (0.0, 0.0347) (64, 0.0333) (112, 0.0323) (256, 0.0279) (512, 0.0157)
    };
    
\addplot[
    color=brown,
    mark=square,
    ]
    coordinates {
    (0.0, 0.0265) (64, 0.02454) (112, 0.0227) (256, 0.01717) (512, 0.00638)
    };
    
\addplot[
    color=red,
    mark=square,
    ]
    coordinates {
    (0.0, 0.01149) (64, 0.0091) (112, 0.00816) (256, 0.0127) (512, 0.0158)
    };
\addplot[
    color=teal,
    mark=square,
    ]
    coordinates {
   (0.0, 0.00199) (64, 0.030) (112, 0.072)
    };
    \legend{350, 500, 750, 1000}

\end{axis}
\end{tikzpicture}
\caption{Increasing inequalities improves results up to a threshold, beyond which it destabilizes the MSE loss, depending on the size of the first hidden layer.}
    \label{fig:mse}
\end{figure}

In Figure \ref{fig:mse} we can see how the MSE evolves in each case 
depending on the number of added inequalities. When the size of the first layer is relatively small, 350 or 500 neurons, the augmenting inequations
produce a considerable impact on recovering the input, lowering the MSE from 0.0347 and 0.0265 to 0.0157 and 0.0063 respectively.

Regarding the efficiency of our attack, for a single sample, the runtime varied between 80 and 400 milliseconds, depending on the number of augmenting equations used. This was measured on an average computer. In the case of a CSP having access to the training process, these times could be substantially reduced by the computational power of the Cloud.

We can see that the augmenting equations are a very particular optimization to our first attack. However, similar optimizations can be possible for other data types, such as sound waves, time-series data (e.g. sensor readings) or geodata (e.g. elevation or pollution maps). For categorical data or text input, this optimization cannot be used.

\subsection{Attack on Convolutional Neural Networks}

We stress that our attack is based on the number of \emph{equations} we can 
simulate the results of the FE decryption. 
In a Convolutional Neural Network (CNN), the first hidden layer usually is very big. For example, in LeNet5 \cite{LeNet5}, the first hidden layer consists of 6 feature \emph{layers}, each of them of size $28\times28 $. This results 
in a number of equations substantially higher than the number of unknown values in the input.
Therefore, in these cases we should be able to completely recover the input image, using the
approach described in Section \ref{attack-large-layer1}. 

However, we made practical tests using a similar approach against CryptoNN applied to LeNet5, by using the linear programming solver used in the previous attack. It should behave the same in the case of a fully 
defined system of equations 
The tests we made were consistent with our expectations, as we were able to fully recover 
the encrypted images. These tests are also available 
in our GitHub repository \cite{github_attack}.

\subsection{Inference Attack}
While the above examples have been presented in the context of training, it is important to emphasize that these vulnerabilities apply also during the inference phase. In this setting, the client encrypts its input and sends it to the server, which uses the functional decryption key to compute the inner product $\langle W, X \rangle$ in plaintext before proceeding with the rest of the computation on unencrypted data. As in the training phase, the attacker has access to both the decrypted inner product $Z_1 = \langle W, X\rangle$ and the weight matrix $W$, enabling identical reconstruction attacks based on this exposed information.

\subsection{A Collusion Attack}

In CryptoNN, the plaintext dataset is encrypted only once, at the pre-processing phase. Then, in the training phase, at each iteration the weights for the neural network change. In a subsequent phase, new decryption keys will be issued for these weights. These new keys can be used to obtain new equations about the encrypted samples. Depending on how many samples are in each iterations and how the blocks on input samples are constructed, it may be possible in CrpytoNN to use decryption keys designated to different blocks with the same sample, in order to obtain more informations.

FENet is encrypting every image with a different public key. This removes the collusion of decryption keys for different samples or blocks of samples: we will not be able to use decryption keys designed for two different input samples on the same input, since their decryption keys do not match.

One minimal way to protect against this attack type would be to shuffle the samples between 
iterations and re-encrypt the dataset. An attacker now needs to reproduce the original 
order using the information leaked from each iteration, and then combine these pieces of information 
to recover the full sample. Note that the information leaked from each iteration
is actually a system of equations, which can be combined regardless of the encryption and decryption keys used to obtain it. 

Since this attack has some obvious counter-measures, we did not implemented it in our work. Our attack is only based on the information we can retrieve in a single run of the forward propagation phase.

\subsection{Detailed Results and Comparison with Previous Attacks}

\subsubsection{Implementation of the attack}

Our attacks are available in our GitHub repository \cite{github_attack}.
We have used the \textbf{SciPy} library \cite{2020SciPy} for linear programming solvers. We have obtained the best results using the "interior-point" solver, despite it being deprecated in newer versions of this software. Since the optimization of the objective function is not applicable in our case, we just placed a vector full of zeroes to the objective function. This means that our solution actually ignores the optimization part and searches for any solution.

\noindent
\bgroup
\setlength\tabcolsep{0.5pt}
\begin{table}
\caption{Comparing the three attacks}
\label{fig:test-no-aug}
    \centering
    \begin{tabular}{|c|c|c||c|}
        \hline 
        E1 & E2 & E3 & Original  \\
        \hline
          {\includegraphics[width=7em]{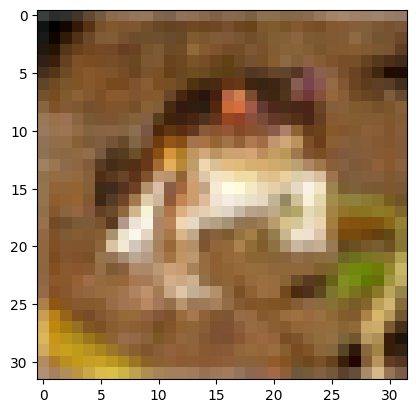} } &
           \includegraphics[width=7em]{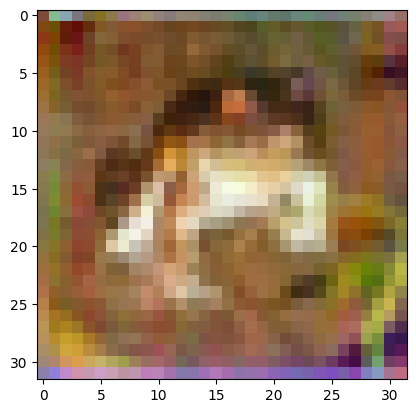} &
          {\includegraphics[width=7em]{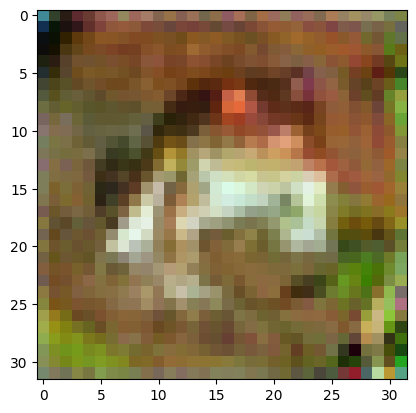} } &
          {\includegraphics[width=7em]{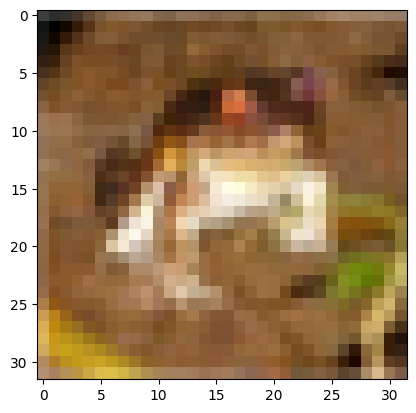} } \\
          \hline
    \end{tabular}
\end{table}
    \egroup

In addition to the experiments conducted over grayscale CIFAR10 using a first dense hidden layer, we have also run experiments against ResNet56 \cite{he2016deep} using unaltered CIFAR10, in order to be able to directly compare our results to DLG\cite{ZhLH2019} and Melis\cite{melis2019exploiting}.

\begin{figure}
    \centering
\begin{tikzpicture}[scale=0.8]

\begin{axis} [ybar, 
ymin=0, ymax=5e-2,
symbolic x coords={Melis, DLG, E1, E2, E3}]
\addplot coordinates {
    (Melis,0.05) 
    (DLG,0.001) 
    (E3,0.0097)
    (E2,0.0093)
    (E1,0.00000000001) 
};
\end{axis}

\end{tikzpicture}

    \caption{Comparison between our attacks (E1, E2, E3), DLG\cite{ZhLH2019} and Melis\cite{melis2019exploiting} on CIFAR Dataset using MSE loss function. While DLG has an error of approximately $10^{-3}$, E1 variant of our experiment fully recovers the image, the MSE being of magnitude $10^-29$, a residual error attributed to floating-point precision.}
    \label{fig:comp}
\end{figure}
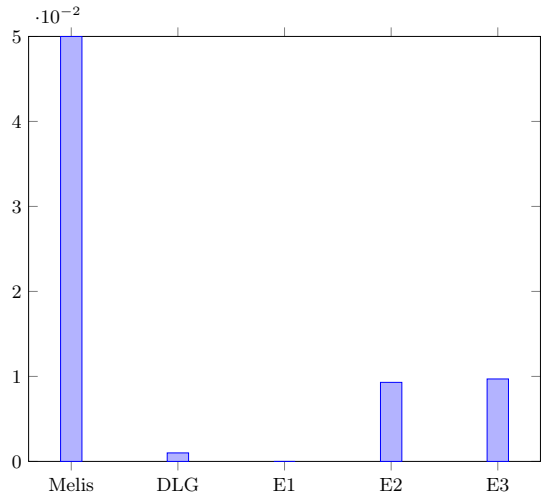

We have run three experiments, as follows:
\begin{itemize}
    \item Experiment 1 (E1). From the 16-channel convolutional first layer of  ResNet56 we have used only 3 channels in order to fully reproduce the image using our first approach - Gaussian elimination. 
    \item Experiment 2 (E2). In order to also test the linear programming approach, we have used only one channel from the first hidden layer. This experiment has no augmentation inequations.
    \item Experiment 3 (E3). This is similar to E2, but it also provides around 60 augmenting inequations. Since we have observed that E2 gives good results especially in the central part of the images, but lascks precision for the margins, we have selected these inequations corresponding to pixels from the margins of the image. 
\end{itemize}

A comparison between our results and the ones obtained in \cite{ZhLH2019}
are summarized in Figure \ref{fig:comp}. While Melis has an error of approximate 0.2, the  
other attack have a considerable lower one, much closer to zero. 
Note that E2 and E3 are reminded here only for theoretical purposes, since we can fully recover the image using the E1 approach. And even there, we are only using 3 out the 16 channles of the convolutional layer used in ResNet56. Therefore, this approach will be able to maintain it's performance even for smaller networks, while DLG will suffer from reducing the number of gradients.  Table \ref{fig:test-no-aug} presents an example on how these three attacks reconstruct an image from CIFAR-10.

Compared to previous attacks, ours is fundamentally different from any other ones operating on privacy-preserving training of neural networks. Table~\ref{tab:attack_comparison} presents a comparative overview of existing neural network data leakage attacks, highlighting the need of features, whether they support single-sample reconstruction, and their associated computational costs.

\begin{table}[hbt]
    \centering
    \caption{Comparison of neural network data leakage attacks in terms of required features, support for single-sample recovery, and computational demands. }
    \label{tab:attack_comparison}
    \begin{tabular}{c||c|c|c}
         Attack &  Features & Single Sample & Computational power \\
         \hline
         \hline
         Carpov\cite{carpov2020illuminating} & Yes & No & Moderate to High \\ 
         Ligier et al. in \cite{ligier2017information} & Yes & No & High \\
          Ours & No & Yes & Low \\
          DLG \cite{ZhLH2019}& No & Yes & High \\
          Melis\cite{melis2019exploiting} & No & Yes & Moderate \\
         
    \end{tabular}
\end{table}

We stress that the main issue addressed by private training of neural networks 
is to protect sensitive data against the cloud server provider. When outsourcing 
the training of a neural network in cloud, it is desirable that the CSP learns as little as possible about the training dataset. However, the service provider is a very capable adversary in this case, having access not only to the dataset, but also to the full description of the neural network and all the weights and activations available in the training process. Hence, our attack only uses the data already available while observing the training process, without interfereing with it (modifying weights for example).
This makes the attack compliant with the Semi Honest adversary, which was the model used in CryptoNN and FENet

\subsection{Mitigation of the Attack}
Our attack succeeds no matter the functional scheme used, but only the idea of having both the inner product and weights in plaintext. 

Hence, a secure training system based on the above-mentioned idea, should not reveal the weights, as the input can be reconstructed. In this section, we are going to present two ways of attack mitigation, one using a more powerful FE scheme and another one that does not use encryption at all but moves a part of the computation on the client’s side.\newline

\subsection{MITIG1 - Solution using FHIPE}
As mentioned before, our attack is possible because the weights are revealed. So, a first idea would be to hide the weights, therefore it came the need of function hiding inner product encryption. However, if the server should now know the weights, it means a trade off in terms of client implication. The client should generate and encrypt the weights and send them to the server along with both the encrypted input and the FHIPE keys for decrypting the inner product. However, the involvement of the client does not stop there, as he has to also calculate the derivative and update the weights locally in order to encrypt them again and send them back to the server.

\begin{figure}
    \centering
\includegraphics[width=0.75\linewidth]{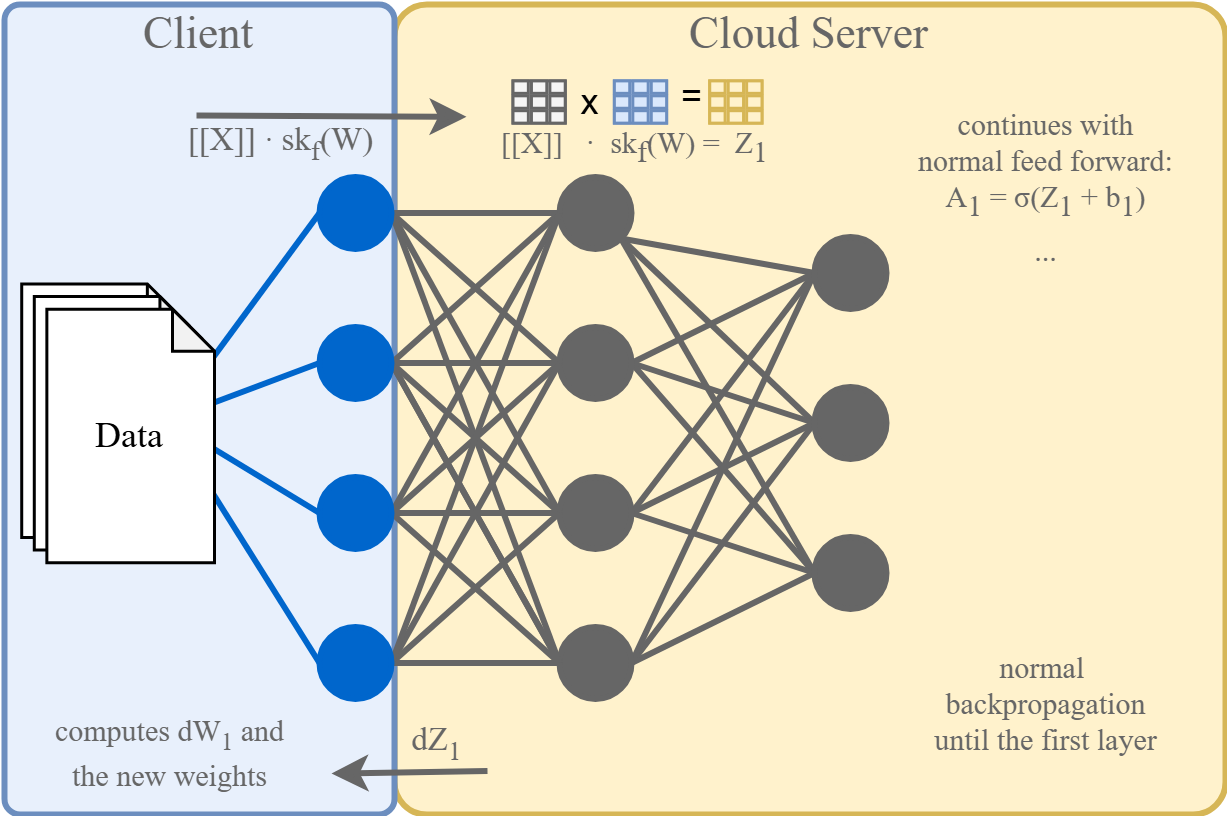}
    \caption{MITIG1 feed-forward and backpropagation using FHIPE. The client generates the weights $W$, encrypts both the input $X$ and weights $W$ using FHIPE, and sends them to the server. Due to the properties of FHIPE, the server computes the inner product $Z_1 = \langle W, X \rangle$ directly in plaintext. During backpropagation, the server sends the derivative of $Z_1$ to the client, who updates and re-encrypts the weights for continued training.}
    \label{fig:mitig1}
\end{figure}

In other words, for a secure training of an NN, a part of the overhead computation has to be done on the client's part. For each feed-forward stage, the client has to generate, encrypt and send the weights to the server. On the other hand, at each backpropagation stage, the client has to calculate the derivative and then update, encrypt and send the new weights. A formal pseudo-code description of MITIG1 is presented in Algorithms \ref{mitig1client} and \ref{mitig1server}. Also, a visual representation of the attack can be consulted in \ref{fig:mitig1}.

The advantage of this system is that the calculation of the inner product is done in the cloud. However, the disadvantages are not easy to overcome. Not only does the client become an integral part of the interaction, but the use of bilinear applications in FHIPE—constructions that perform exceedingly slowly in real-world implementation—further compounds the issue.

Since MITIG1 involves a great amount of computational overhead in the training process, combined with the fact that it also involves an interraction between the clinet and server during the training phase, we proposed another variant of this mitigation, more efficient without losing security.  Therefore, we will focus our attention in the next part on the MITIG2 scheme. However, we stress that MITIG1 still has a great theoretical
importance: although current FE schemes are limited, in the
perspective of a new FE scheme with homomorphic decryption
key capabilities, this scheme could completely remove the
communication overhead, being able to readjust the decryption
key to for the new weights according to the gradients, without
leaking additional information.

\begin{algorithm}[H]
	\caption{MITIG1 Client} 
	\begin{algorithmic}[1]
		\For {$j=1,2,\ldots epoch$}
                \State Generate $W$
			\For {each sample/batch X}
                \State $[[X_i]]=FHIPEEncrypt(X_i)$
                \State $sk_f(W)=FHIPEKeyDerive(W)$
				\State Send $[[X_i]], sk_f(W)$ to server
                \State The server calculates $Z_1 = sk_f(W)\cdot [[X]]$ that continues with feed-forward
                \State On back-propagation, the client receives $dZ_1$
                \State $dW = \frac{1}{m}\cdot dZ_1\cdot x$
                \State $W = W - learningRate \cdot dW$
			\EndFor
		\EndFor
	\end{algorithmic} 
\label{mitig1client}
\end{algorithm}

\begin{algorithm}[H]
	\caption{MITIG1 Server} 
	\begin{algorithmic}[1]
		\For {$j=1,2,\ldots epoch$}
			\For {each sample/batch X}
				\State Receives $[[X_i]], [[W]]$ from the client
                \State Computes $Z_1 = FHIPEDecrypt([[X_i]]\cdot [[W]])$
                \State Feed-forward process
                \State On back-propagation, the server computes $dZ_1$ 
                \State Sends $dZ_1$ to the client
			\EndFor
		\EndFor
	\end{algorithmic} 
\label{mitig1server}
\end{algorithm}

\subsection{MITIG2 - Solution without Encryption}
Building on top of MITIG1, we considered an alternative approach: if the weights are generated on the client side, we could avoid encryption at all. This perspective led to our second concept for a secure system: computing the inner product locally, in plaintext, and sending only the result to the server as input to the first hidden layer. 
This solution is
very similar to Split Learning (SL) \cite{vepakomma2018split} in which the NN is
split in the training phase between a client and a server. 
While usually SL models are focused on distributed and collaborative
learning, we show in the following sections that it can also be
used in our case to provide a strictly stronger privacy guarantee
and better running time compared to previous models using
FE.

For each feed-forward step, we have to recalculate the inner product between the input and the weights. For each backpropagation, we have to calculate the derivative and update the weights. Indeed, the overhead of the client is considerably larger in this case, but it is interesting to see if the overall time is increased because, in fact, we do not work anymore on encrypted data, but on cleartext. Furthermore, the encryption scheme for function-hiding inner product is based on bilinear maps, which are notably slow. Considering the time saved by eliminating the need for encryption, the trade-off of having the client execute some resource-intensive operations on their end no longer appears as burdensome.

\begin{figure}[h]
    \centering

\includegraphics[width=0.75\linewidth]{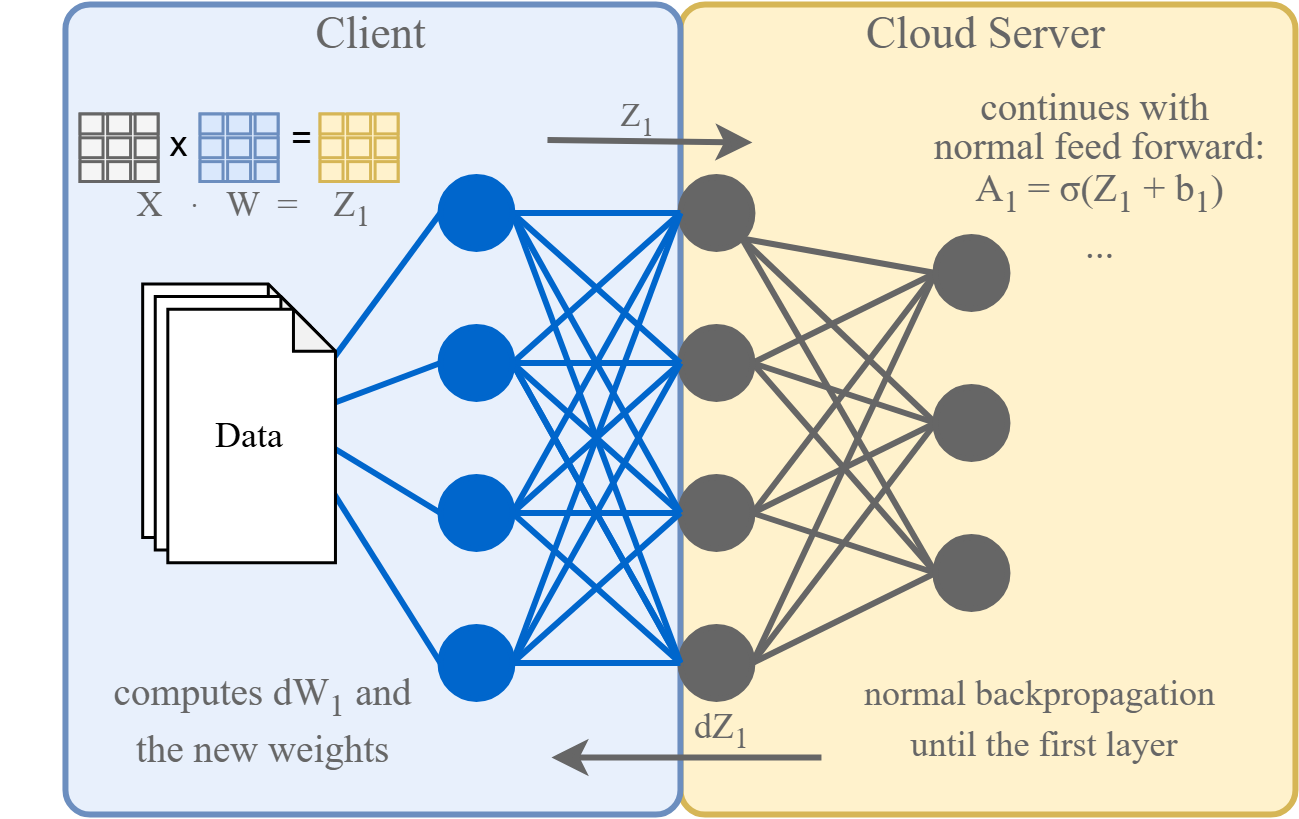}
    \caption{MITIG2 feed-forward and backpropagation without encryption. The client generates the weights and computes $Z_1 = \langle W, X \rangle$, then sends $Z_1$ in plaintext to the server, which continues the standard feed-forward computation. During backpropagation, the server computes the derivative of $Z_1$ and sends it back to the client. The client then updates the weights, recalculates $Z_1$, and the process continues.}
    \label{fig:mitig2}
\end{figure}

In order to speed up the solution, by reducing the interaction between the server and the client, we
could use batch back propagation. In this way, although we will send the same amount of information, we 
will actually receive considerably less information, since the gradients of the input of the first hidden layer will only be sent after a batch is processed. 

After the training phase, the client side will contain the finwl weights for the first layer. Then, these weights can be encrypted using FHIPE, and, then, secure inference can be used. 

\begin{algorithm}
	\caption{MITIG2 Client} 
	\begin{algorithmic}[1]
		\For {$j=1,2,\ldots epoch$}
                \State Generate $W_1$
			\For {each sample/batch X}
				\State Compute $Z_1 = W_1 \cdot X$ 
                \State Send $Z_1$ to the server that continues with feed-forward
                \State On back-propagation, the client receives $dZ_1$
                \State $dW_1 = \frac{1}{m}\cdot dZ_1\cdot X$
                \State $W_1 = W_1 - learningRate \cdot dW_1$
			\EndFor
		\EndFor
	\end{algorithmic} 
\end{algorithm}

\begin{algorithm}
	\caption{MITIG2 Server} 
	\begin{algorithmic}[1]
		\For {each epoch}
			\For {each sample/batch X}
				\State Receives $Z_1$ from the client
                \State $Z_1 \leftarrow Z_1 + b_1$
                \State Feed-forward process on plaintext
                \State On back-propagation, the server computes $dW_2,db_2,dZ_1$ 
                \State Sends $dZ_1$ to the client
			\EndFor
		\EndFor
	\end{algorithmic} 
\end{algorithm}

\subsection{Inference in MITIG1 and MITIG2}
After the secure training phase in both MITIG1 and MITIG2, the weights $W$ 
connecting the input layer to the first hidden layer remain known only to 
the client. To enable secure inference, the client encrypts these weights 
using a functional inner product encryption (FEIP) scheme and sends the 
corresponding functional decryption key $sk_f(W)$ to the server. During 
inference, each client encrypts its input $X$ and transmits the ciphertext 
$[[X]]$ to the server. Using the functional decryption key, the server computes 
the inner product  $Z_1 = sk_f(W)\cdot [[X]]=\langle W,X\rangle$ between the encrypted
input and the encrypted weights, obtaining the result in plaintext. 
From this point onward, the feed-forward process continues entirely in plaintext.

\subsection{Security}

We can observe that our solution provides strictly less information to the cloud server used for training compared to both CryptoNN \cite{xu2019cryptonn} and FENet. The most important part is that it hides the weights between the input and the first hidden layer. 

Considering only one input sample, an attack over this scheme would suppose that given an input $x$ of length $n$, and a matrix of weights $W$ of size $n \times m$, it should be able from knowing only the product $x \cdot W$ to recover both $x$ and $W$. It is obvious that we shrink a much larger quantity of information ($n + nm$ numbers) into a smaller one ($m$ floating point values). This prevents the previously mentioned attack from happening.

On the other hand, while using batch backpropagation, we use multiple samples at the same time during 
the feed forward phase. We will consider the batch size to be $b$. Then, we now have
$b$ input samples, each consisting of $n$ values. These are multiplied with a matrix of weight of size $n\cdot m$, and results in a matrix of size $b \cdot m$. Therefore, having $bn+nm$ unknown values as input, we produce $bm$ values as output. Even though now the number of equations obtained is close to the number of unknowns, the equations are not linear, but quadratic, since both $x$ and $W$ are now unknown to the server. 


\subsection{Comparisons with Other Models}

We have also provided an actual implementation of our MITIG2 scheme, in order to also compare the training time and accuracy of our model, beside the security.
We have used a neural network of a single hidden layer with a dense layer of 300 neurons, against the MNIST Dataset \cite{MNIST}.
Our tests were performed using a 
AMD Ryzen 7 Processor with 32GB of RAM under a Debian operating system.

\bgroup
\setlength\tabcolsep{0.5pt}
\begin{table*}[h]
\caption{Comparing our scheme with similar other ones}
\label{fig:test-dense}
    \centering
    \begin{tabular}{c c  c c}
        \hline 
        Method \qquad & \qquad Training Time \qquad &  \qquad Epoch \qquad & \qquad Accuracy \qquad \\
        \hline 
        \hline  
        unencrypted baseline NN  \cite{panzade2023fenet} & 6.8 s & 2& 96.22\% \\ 
        CryptoNN \cite{xu2019cryptonn} & 52 hrs & 2& 95.49\% \\  
        FENet (FIPE) \cite{panzade2023fenet} & 2 hrs & 2 & 94.87\% \\ 
        FENet (FHIPE) \cite{panzade2023fenet} & 26 hrs & 2 & 94.5\%\\ 
        MITIG2 (unbatched) & 6 min & 2 & 92.6\%\\ 
        MITIG2 (batch 10) & 2 min & 5 & 95.6\%\\ 
        MITIG2 (batch 10) & 5 min & 10 & 96.7\%\\ 
        MITIG2 (batch 100) & 7 min & 30 & 96.8\%\\ 
        \hline
        SecureNN \cite{wagh2019securenn} & 30 hrs & - & 99.15\% \\
        Glyph \cite{lou2020glyph} & 8 days & - & 98.6\%\\

          \hline   \end{tabular}
\end{table*}
    \egroup

As we can see from Table \ref{fig:test-dense}, compared to the other FE-powered NN for training over encrypted data - CryptoNN \cite{xu2019cryptonn} and FENet \cite{panzade2023fenet}, our model outperforms them on both accuracy and training time.

Models relying on secure multi-party computation and fully homomorphic encryption - SecureNN \cite{wagh2019securenn} and Glyph \cite{lou2020glyph} offer a better accuracy, but their training time is considerable higher than ours. 

We have tested our MITIG2 model across various parameters for batch size and epochs, in order
to obtain a better feel about how to accuracy and running time varies with these parameters.
The results can be seen in Table \ref{fig:test-dense-2}.

\bgroup
\setlength\tabcolsep{0.5pt}
\begin{table*}[h]
\caption{Test results for a neural network with a dense first layer}
\label{fig:test-dense-2}
    \centering
    \begin{tabular}{c c  c c}
        \hline 
        Epochs \qquad & \qquad Batch size \qquad &  \qquad Running time \qquad & \qquad Accuracy \qquad \\
        \hline 
        \hline 
        10 & 10 & 5 min & 96.69\% \\  
        10 & 100 & 2 min & 95.51\% \\ 
        10 & 250 & 1 min & 94.62 \%\\ 
        \hline 
        50 & 10 & 21 min & 97.55\% \\  
        50 & 100 & 6 min & 96.8\% \\ 
        50 & 250 & 5 min & 96.36 \%\\ 
        \hline
        25 & 10 & 11 min & 97.39\% \\ 

          \hline   \end{tabular}
\end{table*}
    \egroup

\section{Conclusions}
\label{sec:conclusion}
The current FE approaches have opened a new promising
path for secure training of neural networks over encrypted
data. However, current models are severely limited in terms
of security, leaking a considerable amount of information. We
have shown that under the existing architecture, FE provides
little to no protection during training. Our basic attack, which
requires no input augmentation, can be applied to any type of
training data. The attacks that featured augmentation through
inequalities related to neighboring pixels in the images can be
applied to a limited type of input features, such as images,
sound waves, and videos.

Regarding our proposed mitigation methods, we stress that
they offer strictly better results compared to the previously
mentioned FE-based approaches in terms of security, training
time and accuracy.

Despite information leakage, we consider that the use of a
FE-powered neural network provide a promising foundation
for achieving secure training over encrypted data. We think
that further exploration of FE expressiveness and constructing
new schemes for more powerful functions, such as multi-input
high-degree polynomials, will eventually lead to more secure
neural networks.

\bibliographystyle{alpha}
\bibliography{bibliography}

\section*{Appendix}

\subsection{Failed attack attempt}
\label{app:adam}

Our first thought was guessing the input $X_{guess}$ following the use of \textit{Adam} optimizer in order to minimize the result of $(W_1 \cdot X_{guess} - A_1)^2$. However it did not work because \textit{Adam} tries to find the global minimum, but we could not limit it to give the solution between a certain range. As our testings were made on a classification model where the input was an image, the result should have been a vector with values between $0$ and $255$, the range within a pixel takes values.

\subsection{Functional Encryption}
\label{app:fe} 
\textbf{FE for Basic Operations (FEBO)}

\cite{xu2019cryptonn} et al. presented a FE scheme that supports basic operations.
    \begin{itemize}
        \item \textbf{Setup}: generates $(G,p,g)$, secret $s\leftarrow Z_p$ and returns $(msk, mpk)$ where $msk\leftarrow s$ and $mpk \leftarrow (h,g)$, where $h=g^s$. 
        \item \textbf{KeyDerive} receives 
            \begin{enumerate}
                \item the keys $(mpk,msk)$
                \item the commitment $cmt$
                \item the operation $Delta$
                \item the input of the function $y$
            \end{enumerate} and outputs
            \begin{equation}
                sk_{f_\Delta} = \left \{
              \begin{aligned}
                &cmt^s\dot g^{-y}, && \text{if}\ \Delta=+ \\
                &cmt^s\dot g^{y}, && \text{if}\ \Delta=- \\
                &(cmt^s)^{y}, && \text{if}\ \Delta=* \\
                &(cmt^s)^{y^{-1}}, && \text{if}\ \Delta=/ 
              \end{aligned} \right.
            \end{equation} 
    \end{itemize}

\textbf{FE for Inner Product (FEIP)}

In order to calculate the inner product of two encrypted vectors, \cite{abdalla2015simple} et al. proposed the following FE scheme:
 \begin{itemize}
        \item \textbf{Setup}: generates $(G,p,g)$, secret $(s_1,s_2,\dots s_\eta)\leftarrow Z^\eta_p$ and returns $(msk, mpk)$ where $msk\leftarrow s$ and $mpk \leftarrow (h_i,g)_{i\in |\eta|}$, where $h_i=g^{s_i}$. 
        \item \textbf{Keyderive} receives the secret key $msk$, $y$ and returns $sk_f =\langle y,s/rangle $
        \item  \textbf{Encrypt} takes the public key, the value x to be encrypted, chooses a random $r\leftarrow \mathbf{Z}_p$ and computes 
            \begin{enumerate}
                \item $ct_0=g^r$
                \item $ct_i=h_i^r\dot g^{x_i}$
            \end{enumerate} returning the ciphertext 
        \item \textbf{Decrypt} takes the ciphertext $ct$, the public key $mpk$ and functional key $sk_f$ for the vector $y$ and returns the discrete logarithm $g^{\langle x,y/rangle } =  \prod_{i\in|\eta|}ct_i^{y_i}/ct_0^{sk_f}$
    \end{itemize}

\end{document}